# 3D Gaussian Model for Animation and Texturing

XIANGZHI ERIC WANG* and ZACKARY P. T. SIN*, The Hong Kong Polytechnic University, Hong Kong SAR, China

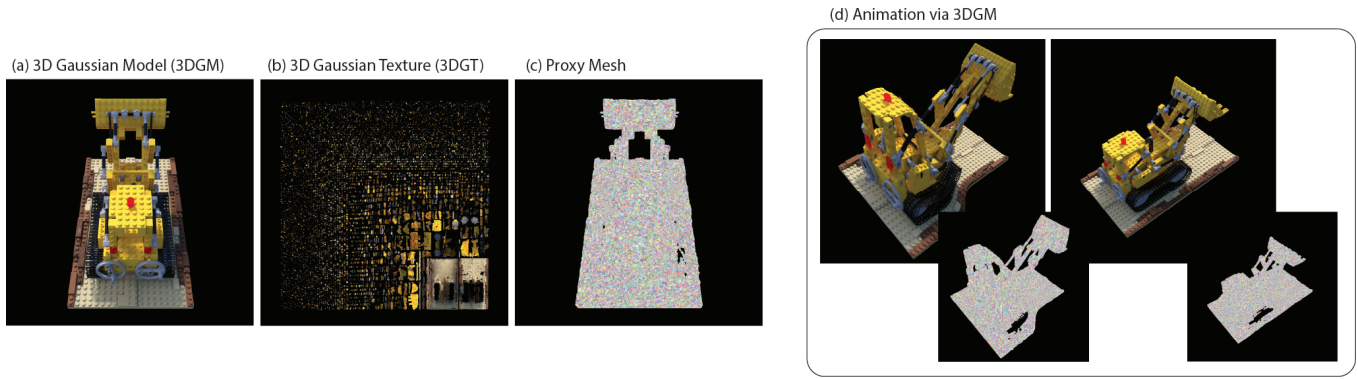

Fig. 1. *(a)* 3D Gaussian Model (3DGM) is analogous to 3D model for 3D Gassian Splatting. By mapping a *(b)* 3D Gaussian Texture (3DGt) to a *(c)* proxy mesh, 3DGM is able to enable *(d)* novel animation and texture transfer; we believe this shows promising potential for our modeling in driving interactive applications.

3D Gaussian Splatting has made a marked impact on neural rendering by achieving impressive fidelity and performance. Despite this achievement, however, it is not readily applicable to developing interactive applications. Real-time applications like XR apps and games require functions such as animation, UV-mapping, and model editing simultaneously manipulated through the usage of a 3D model. We propose a modeling that is analogous to typical 3D models, which we call *3D Gaussian Model* (3DGM); it provides a manipulatable proxy for novel animation and texture transfer. By binding the 3D Gaussians in texture space and re-projecting them back to world space through implicit shell mapping, we show how our 3D modeling can serve as a valid rendering methodology for interactive applications. It is further noted that recently, 3D mesh reconstruction works have been able to produce high-quality mesh for rendering. Our work, on the other hand, only requires an approximated geometry for rendering an object in high fidelity. Application-wise, we will show that our proxy-based 3DGM is capable of driving novel animation without animated training data and texture transferring via UV mapping of the 3D Gaussians. We believe the result indicates the potential of our work for enabling interactive applications for 3D Gaussian Splatting.

## 1 INTRODUCTION

Applying 3D Gaussian Splatting (3DGS) [16] has been a breakthrough methodology for neural rendering [9]. It achieved remarkable fidelity and speed while being relatively robust to different scenes. This novel methodology has already sparked many research works in relation to generative models, digital avatars, and relighting. However, despite these advancements, it seems to be not well-suited for the development of interactive applications such as XR apps and games. Previous works on 3D Gaussians focus on addressing an application problem one at a time, while interactive applications typically require functions such as animation, UV mapping, and model editing simultaneously manipulated through the usage of a 3D model. Thus, how 3D Gaussian Splatting can be used as a rendering method for photo-realistic real-time application development remains an open question.

On the other hand, there is also great progress in mesh reconstruction. Instead of applying neural rendering techniques like 3D Gaussian Splatting, an alternative is to simply use high-fidelity mesh and texturing to enable photo-realistic interactive applications [2, 10, 49]. However, despite the continuous improvement in mesh reconstruction, the reconstructed mesh is not expected to be a perfect replication. Further, it is argued that for many surfaces (with fur, soft bodies, and semi-transparency), simply using mesh for surface modeling may not be sufficient. Hence, it is believed that neural rendering methodologies like 3D Gaussian are likely to have a role to play in surface modeling even as mesh reconstruction works continue to improve.

There is a growing interest in using a proxy in the context of neural rendering; there were previous attempts to map 3D Gaussians to a mesh representation with shell mapping [1], local transform [32], and hybrid representation [10]. A previous work with shell mapping utilizes discretized tetrahedrons; this will lead to C1 discontinuity and rely on self-intersection checking for shell generation which does not work well for all meshes. Another attempt has placed 3D Gaussians within the local spaces of triangles of a mesh; we will show later this approach lacks explicit constraints which will lead to poor deformation results. Last, with a high-fidelity mesh, it has been shown to work well with a hybrid representation; however, it is believed that the surface modeling will rely on the accuracy of the mesh. Further, as mentioned, not all surfaces are suitable to be modeled directly by a mesh.

We propose *3D Gaussian Model* (3DGM), a 3D representation that is analogous to a 3D model for 3D Gaussians Splatting. Similar to other proxy-based work, our work also utilizes a proxy mesh; the key difference is that our work can enable UV mapping without the need for a shell. Instead, our 3D Gaussian Model uses implicit shell mapping, which skips the bijectivity with an injective mapping from texture space to world space. Thus, given a proxy mesh, this enables the 3D Gaussians to be directly densified and optimized in texture space. At the same time, our design also produces a Gaussian texture. Later in our experiments, we will show that the UV-mapped 3D Gaussian Model is able to be manipulated for animation and texturing. We also believe that, by extension, the experiments also show the promise of our work for driving interactive applications. In summary, the following are the contributions of this paper:

- 3D Gaussian Model, a proxy-based representation that is analogous to the 3D model for a typical rasterization-based pipeline
- An injective mapping that enables 3D Gaussians to be optimized in texture space without the need for a shell
- A training strategy that constraints the 3D Gaussians with respect to each triangle, thereby enabling animation and texture mapping

## 2 RELATED WORK

A hot topic in computer graphics, neural rendering [41] refers to a machine-learning-based approach to handling rendering. Usually, work will try to merge the classical computer graphics pipeline with machine learning methodology. Deferred neural rendering [42] and neural radiance field (NeRF) [24] are the early examples of work for neural rendering. Particularly, it seems that recent neural rendering works have a strong connection to NeRF, and most of them are able to produce photo-realistic results for a static scene. 3D Gaussian Splatting (3DGS) [16] is, of course, a recent state-of-the-art NeRF-related model that has made significant advancements for novel view synthesis. Taking the 3D Gaussian as the rendering primitive of the scene, with the splatting technical from [56] and tile-based rendering [3], 3DGS achieves promising novel-view-synthesis results, short training time, and real-time rendering performance.

As far as we know, there is no current work that can simultaneously handle novel animation rendering and texturing in the context of 3DGS. In this section, we will review works from the origin of 3DGS, and its variants focusing on dynamic scenes, scene editing, and avatar reconstructions.

### 2.1 Dynamic Scenes

Prior to 3DGS, there were NeRF-based approaches [24] that could handle dynamic scenes. Works like Nerfies [27, 31] use a neural network to synthesize deformation fields to map across frames. It has also been shown that scene flow [21], the application of a temporal component [46], and representation decomposing [52] also help to control a dynamic scene. More recent 3D-Gaussian-based work on dynamic scenes shows some similarity to these NeRF-based works. For example, like Nerfies, there are 3DGS-driven works that use a neural network to deform the volume [22, 47]. Previous work used a temporal encoder to handle time [45], while a similar effect can also be achieved with a motion factor [18]. There is also an attempt to inject a temporal component such that we have a 4D Gaussian for video reconstruction [20, 48]. This work essentially focuses on modeling a video scene. Thus, they require a pre-determined dynamic scene for training and are likely to lack the capacity to handle novel animation that does not appear in the training data.

On the other hand, there are previous NeRF-based work that can drive animation. Cage-based deformation [12, 28] and shell maps [37] are used previously to enable animated content. Ours is somewhat similar to a previous work [37] in the application of shell mapping, but with our implicit shell mapping, 3DGM do not suffer from discontinuity. In addition, of course, our proposed model also uses 3D Gaussian Splatting as the rendering methodology. There is a recent 3D-Gaussian-based work that also try to address the issue of controllability, but via a neural network [51]. Therefore, it requires animated data for training. Ours, on the other hand, do not require to see any prior animation in order to perform controllable deformation via the proxy.

### 2.2 Content Creation

With the rapid growth of 3DGS, there is also an enthusiastic pursuit of using it for the purpose of content creation. Naturally, coupled with the recent advancement in generative AI, and following the previous success in applying them to NeRF [4, 29], there are already a number of diffusion models [34] that are applied to 3DGS. Foremost, it has been shown that it is possible to use text input to drive the creation of a new 3D object [6], a composition of several objects [43], and subsequently a dynamic object with novel motion [33]. Not limited to objects, but to the generation of a larger scale, there is also a text-based diffusion model that is applied to a 3DGS scene for editing [5], while, similarly, scene generation has also been explored [7, 26]. It has also been shown that an image can be used as a conditional input to control the generation [53]. As the diffusion model usually takes time to generate, there is also an attempt to improve generation speed [50]. It can be seen that it is now possible to use the diffusion model to generate novel content, including motion, for a 3DGS object or scene. Our focus, however, is not on motion synthesis, but on providing a 3D model for 3DGS that can enable generalized motion and texture editing, which will in turn drive interaction applications.

## 2.3 Mesh Reconstruction

While NeRF-based approaches, including 3DGS ones, are able to drive better visual quality, ultimately, they may not provide concrete geometric modeling of the scene. Although that is the case, the scenes provided by NeRF [24], InstantNGP [25], NeuS2 [44] and 3DGS [16] do contain an approximated depth information that can be converted in a 3D mesh reconstruction via techniques such as marching cubes [23]. There are, however, attempts to improve the mesh reconstruction quality from NeRF, for example, by adaptive refinement [39], combining with RGBD signal [2] or coupling the scene understanding with SDF [49]. Empowered by 3DGS, however, SUGAR [10] is able to demonstrate impressive reconstruction results that produce highly detailed mesh. Separately, there is also an attempt to reconstruct a non-rigid mesh via low-ranking coefficient parametrization [8]. However, despite the improved reconstruction quality, it is believed that it is not trivial to produce a mesh reconstruction for all surfaces. Thus, there will always be cases where the mesh reconstruction is not a good approximation to the true surface. Therefore, we believe that our work can strike a balance between a coarse geometric representation and improved surface modeling via a volumetric texture.

## 2.4 Avatar Reconstruction

3DGS provides ample opportunities to researchers to push the boundary for avatar reconstruction. First, there is a work that shows how to capture a human with 3DGS [15]. There are works that used shell maps for avatar reconstruction. The shell is combined with cage-based deformation and neural network to control the placement of the Gaussian [55]. Another work, on the other hand, places layers of generated texture, to model human appearance [1]. Specifically, there is also a work that addresses how to combine the avatar with the surroundings for 3DGS rendering [17]. As can be seen, typical, avatar reconstruction works require a neural network to control the placement of the 3D Gaussian [11, 13]. The neural network is usually needed because of the highly complex nature of human appearance and motion; this limits these methods to data-rich domains like avatar reconstruction. On the other hand, GaussianAvatar (GA)[32] has recently shown that by binding 3D Gaussian locally to triangles, it is possible to bypass the necessity of a neural network. However, it is believed that their method still requires animated data to function. We will later show that the modeling of GaussianAvatar is not suitable for our purpose as it seems that the training video assists the 3D Gaussians in locking locally with the triangles. In contrast to that, the proposed 3DGM do not require animated training data to optimize for handling novel animation, and thus we believe is more suitable to be a foundational 3D model for 3DGS.

## 3 METHODOLOGY

A 3D Gaussian Model, or 3DGM for short, provides a 3D representation that can enable animation and texturing via rendering with 3D Gaussians. The core idea of our approach is to mimic a typical 3D model with mesh and texture. A proxy geometry will essentially serve as the "mesh" while the 3D Gaussians will be placed in the texture space; via UV mapping, the 3D Gaussians will be wrapped around the proxy mesh, forming a 3D Gaussian model that can be used for real-time applications. Some of the key challenges to address are how to handle the nonlinear transformation from texture space to world space and how to constrain the 3D Gaussians with respect to the triangles; solving these issues is the key to enabling a 3D Gaussian model that can simultaneously enable animation and texturing.

### 3.1 Constructing a Proxy Geometry

In the context of the 3DGM, the proxy geometry is essentially the "mesh"; it is used as the basis for UV mapping. In contrast to typical classical 3D models, the proxy mesh does not fully describe the geometric shape of the object. Instead, it is simply a proxy that roughly describes the shape of the object. As long as the proxy geometry provides a sufficient approximation, a 3D Gaussian model will be able to render an object. Hence, similar to deferred neural rendering [42], our works only need a rough proxy for representing objects in a scene and a high-fidelity mesh is not necessary. The proxy can be 3D reconstructed from a structure-from-motion method like COLMAP [36], Instant-NGP [25], and NeuS2 [44], or use marching cubes [23, 38] to approximate the surfaces via a learned 3D Gaussian scene.

### 3.2 3D Gaussian Texture

While a 3DGM is a 3D representation of an object, a 3D Gaussian Texture (3DGT) is essentially a volumetric texture embedded with 3D Gaussians. With this 3D texture, it is possible to use the 3D Gaussians in texture space to describe the surface details on the proxy mesh. Thus, the proxy mesh describes the general shape of the object while the 3DG texture describes the meso-surface details.

Typically, the placement of a 3D Gaussian is described by its position, scale, and rotation. Here, the placement of a 3D Gaussian $\mathcal{G}$ within a 3DG texture is represented differently; it is represented by six bounding points such that we have $\mathcal{G}_u = \{\overleftarrow{u_r}, \overrightarrow{u_r}, \overleftarrow{u_d}, \overrightarrow{u_d}, \overleftarrow{u_f}, \overrightarrow{u_f}\}$ (Figure 2), reminiscent to a bounding box for pose estimation [40]. Each pair of bounding points essentially orientates a basis vector for a 3D Gaussian. For instance, $\overleftarrow{u_r}$ and $\overrightarrow{u_r}$ are orientating the right basis vector $u_r$. We will soon show how the six bounding points can be derived from $\overleftarrow{u_r}, \overrightarrow{u_r}$, a 1D rotation $\theta_d$ and 2D scaling $(s_d, s_f)$, which are the parameters for a 3D Gaussian. It is noted that a point in texture space is $u = (u_u, u_v, u_w)$, where $(u_u, u_v)$ is a 2D point on the texture, while $u_w$ is the "distance" from the 2D plane of the texture. Further, we assume a right-handed coordinate system that uses x-right, y-down, and z-forward.

As mentioned, the bounding points of a 3D Gaussian are derived from the parameters. Foremost, the center of a 3D Gaussian is

$$u_o = \frac{1}{2}(\overleftarrow{u_r} + \overrightarrow{u_r}).$$

Then, the three basis vectors need to be modeled. As mentioned, the rightward basis vector $u_r$ is orientated by $\overrightarrow{u_r}$ and $\overleftarrow{u_r}$; specifically, it is computed by $u_r = (\overrightarrow{u_r} - \overleftarrow{u_r})/|\overrightarrow{u_r} - \overleftarrow{u_r}|$. Based on $u_r$, the other two basis vectors can be derived. With the canonical downward direction $\overline{u}_d = (0, 0, 1)$ in the tangent space of the texture, it can be rotated by $\theta_d$ becoming the downward basis vector $u_d$. Formally, the 1D rotation is done by

$$u_d = u_r \cdot cos(\theta) + (u_r \times \overline{u}_d) \cdot sin(\theta).$$

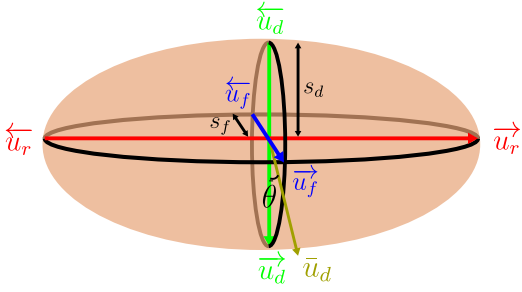

Fig. 2. The placement of a 3D Gaussian in 3DGM is represented by six bounding points, $\overleftarrow{u_r}, \overrightarrow{u_r}, \overleftarrow{u_d}, \overrightarrow{u_d}, \overleftarrow{u_f}$ and $\overrightarrow{u_f}$; they can be derived from the parameters $\overleftarrow{u_r}, \overrightarrow{u_r}, \theta, s_d$ and $s_f$.

With the two basis vectors, the forward basis vector for the 3D Gaussian can be simply computed by

$$u_f = u_r \times u_d.$$

Note that the cross-product for texture space arrange the elements as follows: $u_0 \times u_1 = (u_0, w_0, v_0) \times (u_1, w_1, v_1)$.

While $\overleftarrow{u_r}, \overrightarrow{u_r}$ are parameters on their own, the other four bounding points need to be computed based on two basis vectors $u_d$ and $u_f$, and their related scaling parameters $s_d$ and $s_f$. Simply, the four bounding points are computed with

$$\overleftarrow{u_d} = u_o - s_d \cdot u_d, \overrightarrow{u_d} = u_o + s_d \cdot u_d,$$
$$\overleftarrow{u_f} = u_o - s_f \cdot u_f \text{ and } \overrightarrow{u_f} = u_o + s_f \cdot u_f.$$

In total, the six bounding points determine where a 3D Gaussian is at in the texture space.

### 3.3 Implicit Shell Mapping

A typical way to apply a volumetric texture on a mesh is to use shell mapping [30]. Most often, the mapping will require breaking extruded prisms into tetrahedrons for Barycentric correspondence, but this discretization will lead to C1 discontinuity. The other method is to handle the non-linearity of a prism by segmenting it for numerical approximation [14]. Either way, the construction of the shell is not trivial. The thickness of the shell needs to be somehow determined, and this is usually done manually. Furthermore, as most mesh will have concavity, the constructed shell will require self-intersection checking to avoid the prisms overlapping with one another. By preventing overlapping, the shell will have uneven thickness.

With 3DGM, we propose implicit shell mapping that can avoid the complexity of explicitly constructing a shell. With a UV-mapped mesh, a triangle $\Delta ABC$ will have vertices $v_A, v_B$ amd $v_C$ mapped to UV coordinates $u_A, u_B$ and $u_C$. Any point in the UV space $u$ can be converted to the Barycentric coordinate of the triangle $\phi$ by comparing areas such that

$$\phi = (\phi_A, \phi_B, \phi_C) = \Phi_{u \to \phi}(u_u, u_v; u_A, u_B, u_C),$$

where $\phi_A$, $\phi_B$ and $\phi_C$ are the elements of $\phi$.

For the w-component of a UV coordinate $u_w$, an issue exists in that it has no connection with the physical properties of the 2D $u_u$ and $u_v$. The latter two components are used to map to the triangles of the mesh and thus are related to the size of the triangle. $u_w$, on the other hand, has no relation to the triangle itself. It has been found that applying a similar learning rate of $u_u$ and $u_v$ on $u_w$ will lead to optimization instability due to the 1D rotation parameter $\theta_d$. As such, we calculate a w-component scaler for each triangle

$$\omega_\Delta = \frac{1}{3}(\frac{v_A - v_B}{v_A - v_B} + \frac{v_A - v_C}{v_A - v_C} + \frac{v_B - v_C}{v_B - v_C}),$$

which can in turn become a scaler for each vertex

$$\omega_v = \frac{1}{|v_\Delta|} \sum_\delta^{|v_\Delta|} \omega_\delta,$$

where $v_\Delta$ is the set of triangles that are connected with the vertex $v$. With the scaler, an adjusted w-component can be computed with the following

$$\phi_w = \phi_A \omega_A + \phi_B \omega_B + \phi_C \omega_C.$$

With the above, the Barycentric coordinate can be further transformed into world space with

$$v = \Phi_{\phi \to v}(\phi) = \phi_A v_A + \phi_B v_B + \phi_C v_C + \phi_w(\phi_A n_A + \phi_B n_B + \phi_C n_C), \quad (1)$$

where $n_A$, $n_B$ and $n_C$ are the vertex normal of the triangle $\Delta ABC$.

With the above, we can see that each bounding point of the Gaussian can be projected into world space, for instance $\overleftarrow{v_r} = (\Phi_{\phi \to v} \circ \Phi_{u \to \phi})(\overleftarrow{u_r})$. We will soon show how the bounding points in world space can be used to project the 3D Gaussian into world space as well.

Although it can enable C1 continuity without numerical approximation, the downside of implicit shell mapping is that the bijectivity of typical shell mapping is lost. Instead, the mapping can only be done unidirectional from texture space to world space.

### 3.4 Sheared 3D Gaussian

Shell mapping, whether uses our implicit design or not, needs to consider non-linearity. In our case, the non-linearity is introduced by Equation 1. Thus, the linear relationships between the six bounding points cannot be maintained when projecting from texture space to world space. Due to this reason, we propose a sheared 3D Gaussian. Its covariance matrix $\Sigma$ is computed by a shearing matrix such that

$$\Sigma = Sh \cdot Sh^T, Sh = \begin{bmatrix} \overrightarrow{v_r} - v_o \\ \overrightarrow{v_d} - v_o \\ \overrightarrow{v_f} - v_o \end{bmatrix}^T.$$

Again, although implicit shell mapping does not enforce bijectivity, injective mapping still enables backpropagation. Thus, the sheared 3D Gaussian can provide a gradient to update the parameters in texture space.

### 3.5 Barycentric Constraints

It is important to constrain a 3D Gaussian within the triangle it is associated with. Each triangle of the proxy mesh will have 3D Gaussians to model its local surface. In order to ensure that a 3D Gaussian stays with the implicit prism of the triangle, we have introduced a Barycentric regularizer that penalizes 3D Gaussians that are (partially) outside of the implicit prism. This regularizer is applied to all six bounding points of a 3D Gaussian, and we first check if any of their Barycentric coordinate components is smaller than 0. If that is the case, we first compute a reference coordinate $\hat{\phi}$, which specifies the closest point in Barycentric space such that the bounding point is not outside of the triangle:

$$\hat{\phi} = \frac{1-t}{3} + t\phi, \ t = (1 - 3 \cdot \min(\phi_A, \phi_B, \phi_C))^{-1}$$

This point is then projected to world space to become a reference on where the 3D Gaussian bounding point should be in world space. As such, we can now calculate the regularizer for a bounding point:

$$\mathcal{L}_\phi(\mathcal{G}) = \frac{1}{6} \sum_{\phi \in \mathcal{G}_\phi} \text{ReLU}(||\Phi_{\phi \to v}(\hat{\phi}) - \Phi_{\phi \to v}(\phi)||_2^2 - \epsilon_\phi),$$

where $\mathcal{G}_\phi$ is the set of the bounding points' Barycentric coordinate and $\epsilon_\phi$ is a hyperparameter that allows a bounding point to slightly go outside the triangle. Empirically, we set $\epsilon_\phi = 0.01$.

Further, to avoid the 3D Gaussians from straying too far away from the triangle, we also introduce an extrusion regularizer which penalizes with the following:

$$\mathcal{L}_w(\mathcal{G}) = \sum_{\phi_w \in \mathcal{G}_{\phi_w}} |\phi_w|$$

where $\mathcal{G}_{\phi_w}$ is the set of the bounding points' w-component Barycentric coordinate.

### 3.6 Initialization

For each triangle, three initializing 3D Gaussian are assigned to it. Specifically, given triangle $\Delta ABC$, the starting 3D Gaussian are placed such that $\overleftarrow{u_r}^{(0)} = u_A, \overleftarrow{u_r}^{(1)} = u_B, \overleftarrow{u_r}^{(2)} = u_C$ and $\overrightarrow{u_r}^{(0)} = \overrightarrow{u_r}^{(1)} = \overrightarrow{u_r}^{(2)} = (u_A + u_B + u_C)/3$, where $\overleftarrow{u_r}^{(0)}$ and $\overrightarrow{u_r}^{(0)}$ is the bounding points for the first initializing 3D Gaussian. With densification, new 3D Gaussians will be split or cloned from an existing 3D Gaussian. The new 3D Gaussians will remain in the same triangle as the existing 3D Gaussians.

### 3.7 Optimization

The optimization function of 3DGM is similar to the original 3DGS. The main addition is the Barycentric regularizer and the extrusion regularizer discussed in . The loss is as follows:

$$\mathcal{L} = \lambda_1 \mathcal{L}_1 + (1 - \lambda_1)\mathcal{L}_{D-SSIM} + \sum_{\mathcal{G} \in M_\mathcal{G}} \lambda_\phi \mathcal{L}_\phi(\mathcal{G}) + \lambda_w \mathcal{L}_w(\mathcal{G}),$$

where $M_\mathcal{G}$ is the set of 3D Gaussians on the 3DGM. For the hyperparameters, we set $\lambda_1 = 0.8$, $\lambda_\phi = 10^5$ and $\lambda_w = 1$.

## 4 RESULT AND EVALUATION

The goal of 3DGM is to be a 3D model equivalent for 3DGS, which will be able to drive interactive applications. Here, we showcase 3DGM's ability to support novel animation and texture transfer. First, we present the quantitative result of 3DGS in handling novel animation, with an ablation study on different proxy mesh inputs. Then, we showcase the 3D Gaussian Texture (3DGT) and subsequently, how to perform texture transfer with it.

### 4.1 Animation

A 3D model is constructed in its static form and can later be used to be deformed for novel animation that is separate from the surface information. This assumption is also useful for interactive application development as the usual process of building animated content is that new animation can be added based on the static 3D model. Thus, with this in mind, the models that are evaluated here have not seen any animated training data before. Instead, their surface information is learned from the different views of a static object. After the training, the animation can be applied to the 3D Gaussianlized object by animating the proxy mesh, and the Gaussians move, stretch, and rotate based on the transformation of their binding triangles in an automatic manner. In this way, one with the knowledge of animating a mesh could create animation on such an object without touching the particularized individual 3D Gaussians.

To have a relatively comprehensive quantified result of 3DGM, we trained the original 3D Gaussian model, the GaussianAvatar (GA) [32] with our implemented regularizers, 3DGM without the $\mathcal{L}_\phi$, and 3DGM. We adopted three scenes, the Chair and Lego from the NeRF [24] synthetic dataset, and the Stanford Armadillo [19]. To prepare the proxy meshes, we used NeuS2 [44] for Lego and Chairs. For the Armadillo's proxy, we used Blender to decimate the mesh by a factor of 0.5. The three scenes are further enriched with animations and provide object-animated ground truth images for calculating the metrics. Thus, all the models are testified with both static and animated status, except 3DGS which does not support novel animation. For the metrics, we used the standard LPSIS [54], SSIM, and PSNR to compare the fidelity of our animated rendering. We also introduced the intersection over union (IoU) to compare with as we would like to specifically evaluate whether a model can retain the shape of the object during animation.

To the best of our knowledge, given a lack of animated training data, we are one of the earliest works that can handle novel animation. As mentioned, it is possible to use a high-fidelity mesh to replicate photo-realistic results. We, however, argue that mesh reconstruction works that produce high-fidelity mesh will result in a loss of surface detail. Thus, we did not compare them at this stage. However, GA [32] has shown how to embed a 3D Gaussian into the local space of a triangle. Its key strategy is to initialize 3D Gaussians on the proxy mesh triangles and enforce a scaling and position regularizer to keep the 3D Gaussian near the triangles in question. Although their work shows similarity to our work, it is believed that their model has not considered the non-linearity that occurs during animation. Instead, they rely on animated training data to improve the robustness of their model in handling non-linearity introduced by novel animation. Thus, 3DGM addressed an important question of how to enable 3D modeling for 3DGS as it does not require animated training data to handle novel animation. To the best of our ability, we have implemented the regularizers from GA to bind the 3D Gaussians to the triangles.

As shown on Figure 5, Figure 6, Figure 7, Figure 8 and Table 1, our full model is able to have the overall best fidelity compared to others, in both static novel view synthesis and novel animation synthesis. For the static evaluation, 3GM is comparable to 3DGS while even achieving better PSNR in Chair and Lego scenes. On the other hand, 3GM has the best SSIM and LPIPS results and slightly lower scores on the PSNR and IoU compared to GA [32] in the context of novel animation. The result also shows the importance of our Barycentric regularizer, without it, the result is somewhat similar to GA's. In general, it seems that the result shows the 3DGM is able to render for novel animation.

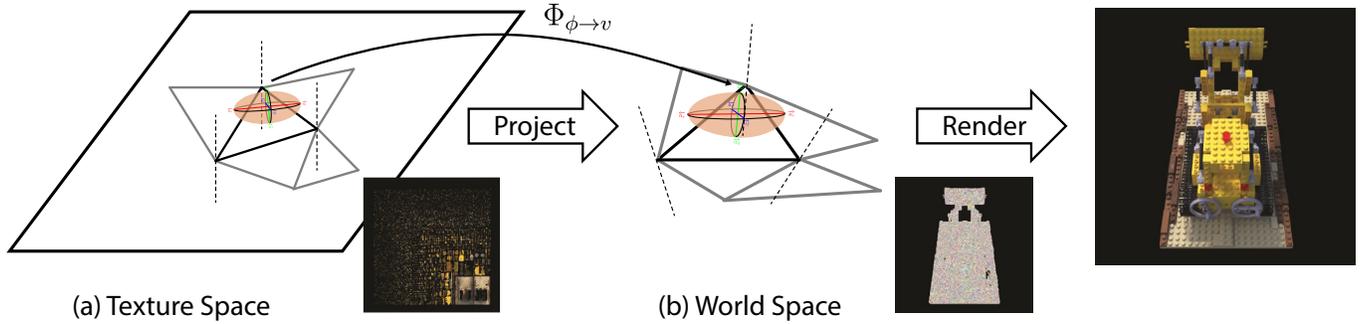

Fig. 3. An overview on how a 3D Gaussian Model renders. First, 3D Gaussians are stored on a 3D Gaussian Texture. They are projected into world space via barycentric correspondence. As the shell space is not linear, the six bounding points are used for projection, which models a sheared 3D Gaussian. This sheared 3D Gaussian is used for rendering.

To illustrate more about the fidelity of 3GM in the wide acceptation proxy mesh input of our model, we performed an ablation study toward the Lego scene which has the most complex geometry among the scenes, using different resolutions of the proxy to train 3DG and inference the models with both static and animated formats. Table 2 presents the metrics on full 3DG trained upon resolutions of 100% (380k faces), 40% (150k), 20% (76k), and 10% (38k) proxy meshes. The results show that the performance only declined slightly with the massive decrease in the face numbers. As a rule of thumb, a high-resolution mesh is typically harder to re-construct or reconstruct, as well as more difficult for skeleton binding and driven by animation. The wide acceptation of proxy meshes can infer the cost-efficiency of running 3DG. The render result of the Lego scene with different resolutions (Figure 9) demonstrated the capability of 3DGM to accept multiple resolutions of proxy meshes further. From fine to coarse proxies, the appearance of the sub-figures mostly looks the same.

### 4.2 Texture Transfer

Another application of 3DGM is the ability to produce a texture of 3D Gaussians (3DGT). It is believed this will assist in handling 3D modeling work for 3DGS such as re-texturing, which will enable the same texture to be applied to other objects.

Here, we used surface mapping [35] to synchronize the UV mapping between different objects in order to enable texture transfer. The transfer result can be seen in Figure 4. Note that we used the same texture and applied it to objects with different shapes. As shown in the figure, 3DGM is able to transfer a texture from one object to another. The texture originally comes from a furry cartoon cow. It is then transferred to a more realistically modeled horse, camel, and cow. Despite the difference in geometry, we can see the appearance and surface detail have been transferred to the target objects. This result shows that the proxy mesh and 3DGT are separated, similar to a typical 3D model with mesh and (albedo) textures. Thus, 3DGM can enable us to capture an object in the real world and extract its texture. With a similarly shaped object, we can transfer the surface appearance from the captured object to the new object.

## 5 LIMITATION AND FUTURE WORK

As shown in the result, our model is robust to the coarseness of the proxy geometry. However, ultimately, 3DGM still requires an approximated geometry to represent a target. If such an approximating mesh cannot be produced, our model will not be able to model the surface in question. As such, 3DGM is limited by the proxy mesh. In our experiments, as we want to test 3DGM's ability to handle a coarse geometry, we have used mesh constructed via NeurS2 [44] and marching cubes. As such, in the future, to evaluate the best fidelity of the current 3DGM, we will conduct experiments with the help of state of the art mesh reconstruction algorithm. We believe 3DGM will be able to produce better image quality while enabling animation and texture transfer. Further, there will be cases where the object may not be suitably represented by a mesh (furry surface), or unable to be reconstructed as a mesh (e.g. vegetation, very thin objects). It is suggested that investigating other forms of primitives to attach the 3D Gaussians can be an interesting future direction to explore.

## 6 CONCLUSION

We present 3D Gaussian Model (3DGM), a surface modeling method that enables 3D Gassian Splatting (3DGS) to drive interactive applications. It is analogous to a 3D model for 3DGS with a proxy mesh and a texture. The proxy mesh is an approximated geometry of the target object while the 3D Gaussian Texture (3DGT) is a volumetric texture embedded with 3D Gaussian to model the surface information. To ensure the 3D Gaussians remain as useful surface modeling to the proxy, we have utilized an implicit shell mapping design. The six bounding points of the 3D Gaussians can be projected from texture space to world space. Due to non-linearity, the 3D Gaussians in world space are sheared. To further constrain the 3D Gaussians to the paired triangles, we have also introduced a Barcentric regularizer.

In our experiments, we have shown that 3DGM is able to handle animation and texturing; to the best of our knowledge, we are the first to show how to drive novel animation and texturing simultaneously for 3DGS. We believe that our work lays the foundation for how to utilize 3DGS for developing real-time applications such as XR apps and games.

Table 1. Quantitative evaluation of our method in static and driving animation.

| | Chair | | | | Lego | | | | Armadillo | | | |
|---|---|---|---|---|---|---|---|---|---|---|---|---|
| **(a) Static** | PSNR ↑ | SSIM ↑ | LPIPS ↓ | IoU ↑ | PSNR ↑ | SSIM ↑ | LPIPS ↓ | IoU ↑ | PSNR ↑ | SSIM ↑ | LPIPS ↓ | IoU ↑ |
| 3DGS [16] | 35.29 | 0.988 | 0.009 | 0.995 | 35.84 | 0.982 | 0.013 | 0.983 | 47.29 | 0.998 | 0.001 | 0.995 |
| GA [32] Regularizers | 35.01 | 0.987 | 0.010 | 0.994 | 35.24 | 0.979 | 0.015 | 0.989 | 46.74 | 0.998 | 0.001 | 0.995 |
| 3DGM (Ours w/o $\mathcal{L}_\phi$) | 35.44 | 0.988 | 0.009 | 0.995 | 35.74 | 0.982 | 0.013 | 0.989 | 47.77 | 0.999 | 0.001 | 0.995 |
| 3DGM (Ours) | 35.45 | 0.988 | 0.008 | 0.995 | 35.73 | 0.982 | 0.013 | 0.991 | 47.78 | 0.999 | 0.001 | 0.995 |
| **(b) Animated** | | | | | | | | | | | | |
| GA [32] Regularizers | 30.67 | 0.980 | 0.013 | 0.988 | 27.21 | 0.920 | 0.070 | 0.978 | 34.28 | 0.983 | 0.015 | 0.989 |
| 3DGM (Ours w/o $\mathcal{L}_\phi$) | 30.60 | 0.982 | 0.010 | 0.987 | 27.84 | 0.941 | 0.044 | 0.985 | 32.44 | 0.974 | 0.022 | 0.981 |
| 3DGM (Ours) | 30.63 | 0.982 | 0.010 | 0.987 | 28.25 | 0.945 | 0.040 | 0.987 | 36.27 | 0.987 | 0.011 | 0.992 |

Table 2. Ablation study of different resolution proxy meshes on the Lego scene

| | Static | | | | Animated | | | |
|---|---|---|---|---|---|---|---|---|
| **Resolution** | PSNR ↑ | SSIM ↑ | LPIPS ↓ | IoU ↑ | PSNR ↑ | SSIM ↑ | LPIPS ↓ | IoU ↑ |
| 100% | 35.73 | 0.982 | 0.013 | 0.991 | 28.25 | 0.945 | 0.040 | 0.987 |
| 40% | 35.57 | 0.981 | 0.014 | 0.989 | 27.99 | 0.940 | 0.045 | 0.983 |
| 20% | 35.50 | 0.981 | 0.014 | 0.987 | 27.77 | 0.937 | 0.048 | 0.981 |
| 10% | 35.29 | 0.980 | 0.015 | 0.986 | 27.51 | 0.933 | 0.052 | 0.979 |

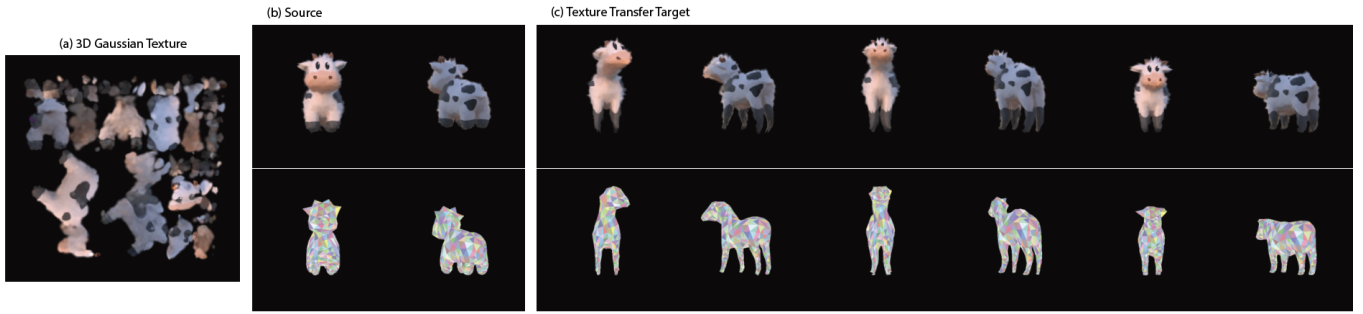

Fig. 4. Via the 3D Gaussian Texture (3DGT), 3D Gaussian Model (3DGM) enables transfer of surface appearance from source object to target object.

REFERENCES
[1] Rameen Abdal, Wang Yifan, Zifan Shi, Yinghao Xu, Ryan Po, Zhengfei Kuang, Qifeng Chen, Dit-Yan Yeung, and Gordon Wetzstein. 2023. Gaussian Shell Maps for Efficient 3D Human Generation. arXiv:2311.17857 [cs.CV]
[2] Dejan Azinović, Ricardo Martin-Brualla, Dan B Goldman, Matthias Nießner, and Justus Thies. 2022. Neural rgb-d surface reconstruction. In *Proceedings of the IEEE/CVF Conference on Computer Vision and Pattern Recognition*. 6290–6301.
[3] WFPW Burgers. 2005. Tile-Based Rendering. *traffic* 600 (2005), 3.
[4] Hansheng Chen, Jiatao Gu, Anpei Chen, Wei Tian, Zhuowen Tu, Lingjie Liu, and Hao Su. 2023. Single-Stage Diffusion NeRF: A Unified Approach to 3D Generation and Reconstruction. In *ICCV*.
[5] Yiwen Chen, Zilong Chen, Chi Zhang, Feng Wang, Xiaofeng Yang, Yikai Wang, Zhongang Cai, Lei Yang, Huaping Liu, and Guosheng Lin. 2023. GaussianEditor: Swift and Controllable 3D Editing with Gaussian Splatting. arXiv:2311.14521 [cs.CV]
[6] Zilong Chen, Feng Wang, and Huaping Liu. 2023. Text-to-3d using gaussian splatting. *arXiv preprint arXiv:2309.16585* (2023).
[7] Jaeyoung Chung, Suyoung Lee, Hyeongjin Nam, Jaerin Lee, and Kyoung Mu Lee. 2023. LucidDreamer: Domain-free Generation of 3D Gaussian Splatting Scenes. *arXiv preprint arXiv:2311.13384* (2023).
[8] Devikalyan Das, Christopher Wewer, Raza Yunus, Eddy Ilg, and Jan Eric Lenssen. 2023. Neural Parametric Gaussians for Monocular Non-Rigid Object Reconstruction. *arXiv preprint arXiv:2312.01196* (2023).
[9] Lin Yen-Chen Frank Dellaert. 2021. Neural Volume Rendering: NeRF And Beyond. *arXiv preprint arXiv:2101.05204* (2021).
[10] Antoine Guédon and Vincent Lepetit. 2023. SuGaR: Surface-Aligned Gaussian Splatting for Efficient 3D Mesh Reconstruction and High-Quality Mesh Rendering. *arXiv preprint arXiv:2311.12775* (2023).
[11] Shoukang Hu and Ziwei Liu. 2023. GauHuman: Articulated Gaussian Splatting for Real-Time 3D Human Rendering. *arXiv preprint* (2023).
[12] Clément Jambon, Bernhard Kerbl, Georgios Kopanas, Stavros Diolatzis, George Drettakis, and Thomas Leimkühler. 2023. NeRFshop: Interactive Editing of Neural Radiance Fields. *Proceedings of the ACM on Computer Graphics and Interactive Techniques* 6, 1 (2023).

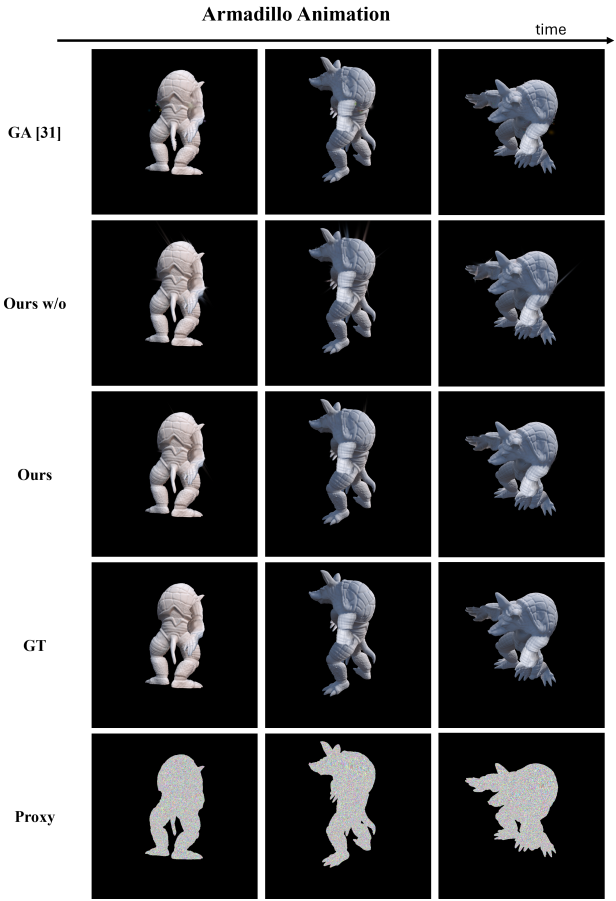

Fig. 5. The Armadillo animated with skeletal deformation.

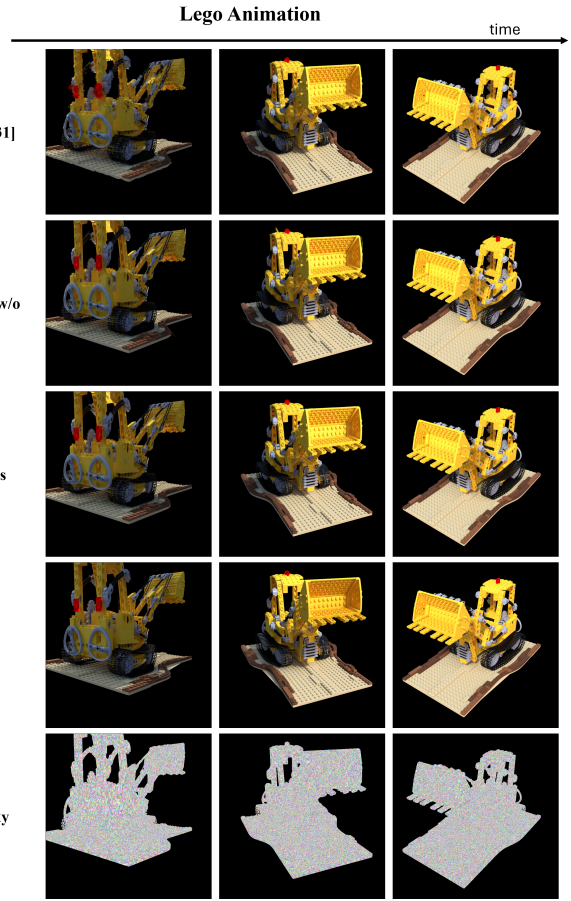

Fig. 6. The Lego animated with lattice deformation.

[13] Rohit Jena, Ganesh Subramanian Iyer, Siddharth Choudhary, Brandon Smith, Pratik Chaudhari, and James Gee. 2023. SplatArmor: Articulated Gaussian splatting for animatable humans from monocular RGB videos. *arXiv preprint arXiv:2311.10812* (2023).

[14] Stefan Jeschke, Stephan Mantler, and Michael Wimmer. 2007. Interactive smooth and curved shell mapping. In *Proceedings of the 18th Eurographics conference on Rendering Techniques*. 351–360.

[15] HyunJun Jung, Nikolas Brasch, Jifei Song, Eduardo Perez-Pellitero, Yiren Zhou, Zhihao Li, Nassir Navab, and Benjamin Busam. 2023. Deformable 3D Gaussian Splatting for Animatable Human Avatars. *arXiv preprint arXiv:2312.15059* (2023).

[16] Bernhard Kerbl, Georgios Kopanas, Thomas Leimkühler, and George Drettakis. 2023. 3D Gaussian Splatting for Real-Time Radiance Field Rendering. *ACM Transactions on Graphics* 42, 4 (July 2023). https://repo-sam.inria.fr/fungraph/3d-gaussian-splatting/

[17] Muhammed Kocabas, Jen-Hao Rick Chang, James Gabriel, Oncel Tuzel, and Anurag Ranjan. 2023. Hugs: Human gaussian splats. *arXiv preprint arXiv:2311.17910* (2023).

[18] Agelos Kratimenos, Jiahui Lei, and Kostas Daniilidis. 2023. DynMF: Neural Motion Factorization for Real-time Dynamic View Synthesis with 3D Gaussian Splatting. *arXiv preprint arXiv:2312.00112* (2023).

[19] Venkat Krishnamurthy and Marc Levoy. 1996. Fitting smooth surfaces to dense polygon meshes. In *Proceedings of the 23rd annual conference on Computer graphics and interactive techniques*. 313–324.

[20] Zhan Li, Zhang Chen, Zhong Li, and Yi Xu. 2023. Spacetime Gaussian Feature Splatting for Real-Time Dynamic View Synthesis. *arXiv preprint arXiv:2312.16812* (2023).

[21] Zhengqi Li, Simon Niklaus, Noah Snavely, and Oliver Wang. 2021. Neural Scene Flow Fields for Space-Time View Synthesis of Dynamic Scenes. 6498–6508.

[22] Yiqing Liang, Numair Khan, Zhengqin Li, Thu Nguyen-Phuoc, Douglas Lanman, James Tompkin, and Lei Xiao. 2023. GauFRe: Gaussian Deformation Fields for Real-time Dynamic Novel View Synthesis. *arXiv preprint arXiv:2312.11458* (2023).

[23] William E. Lorensen and Harvey E. Cline. 1987. Marching cubes: A high resolution 3D surface construction algorithm. In *Annual Conference on Computer Graphics and Interactive Techniques (SIGGRAPH) (SIGGRAPH '87)*. Association for Computing Machinery, New York, NY, USA, 163–169. https://doi.org/10.1145/37401.37422

[24] Ben Mildenhall, Pratul P Srinivasan, Matthew Tancik, Jonathan T Barron, Ravi Ramamoorthi, and Ren Ng. 2021. Nerf: Representing scenes as neural radiance fields for view synthesis. *Commun. ACM* 65, 1 (2021), 99–106.

[25] Thomas Müller, Alex Evans, Christoph Schied, and Alexander Keller. 2022. Instant Neural Graphics Primitives with a Multiresolution Hash Encoding. *ACM Trans. Graph.* 41, 4, Article 102 (July 2022), 15 pages. https://doi.org/10.1145/3528223.3530127

[26] Hao Ouyang, Tiancheng Sun, Stephen Lombardi, and Kathryn Heal. 2023. Text2Immersion: Generative Immersive Scene with 3D Gaussians. *Arxiv* (2023).

[27] Keunhong Park, Utkarsh Sinha, Jonathan T Barron, Sofien Bouaziz, Dan B Goldman, Steven M Seitz, and Ricardo Martin-Brualla. 2021. Nerfies: Deformable neural radiance fields. 5865–5874.

[28] Yicong Peng, Yichao Yan, Shengqi Liu, Yuhao Cheng, Shanyan Guan, Bowen Pan, Guangtao Zhai, and Xiaokang Yang. 2022. CageNeRF: Cage-based neural radiance field for generalized 3D deformation and animation. *Advances in Neural Information Processing Systems* 35 (2022), 31402–31415.

[29] Ben Poole, Ajay Jain, Jonathan T. Barron, and Ben Mildenhall. 2022. DreamFusion: Text-to-3D using 2D Diffusion. *arXiv* (2022).

[30] Serban D Porumbescu, Brian Budge, Louis Feng, and Kenneth I Joy. 2005. Shell maps. *ACM Transactions on Graphics (TOG)* 24, 3 (2005), 626–633.

[31] Albert Pumarola, Enric Corona, Gerard Pons-Moll, and Francesc Moreno-Noguer. 2020. D-NeRF: Neural Radiance Fields for Dynamic Scenes. 10318–10327.

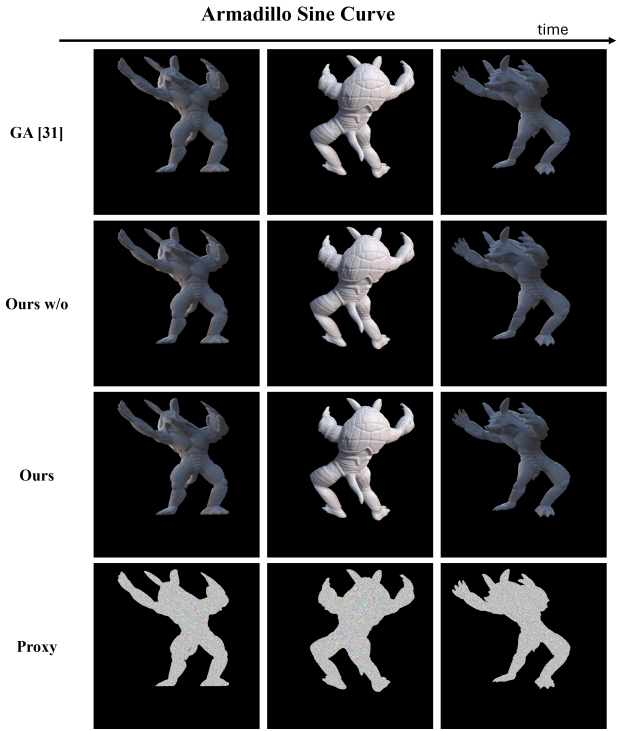

Fig. 7. The Armadillo animated with a sine curve.

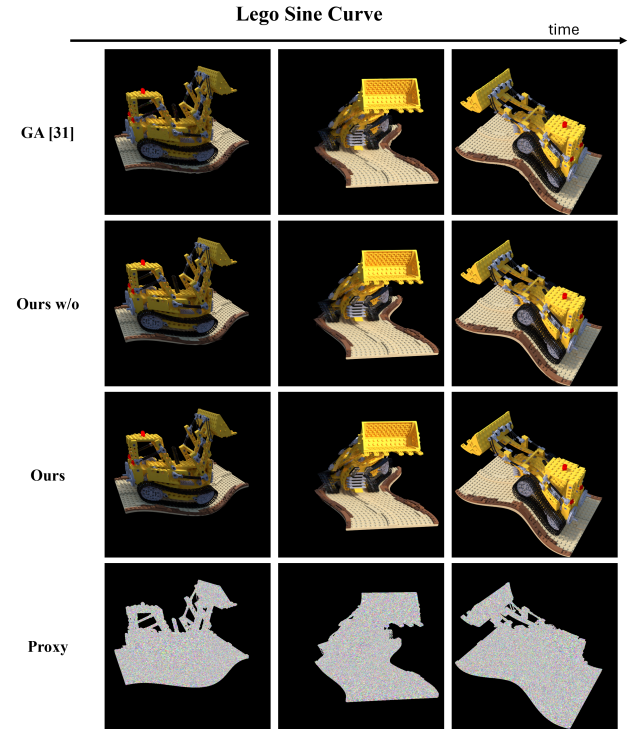

Fig. 8. The Lego animated with a sine curve.

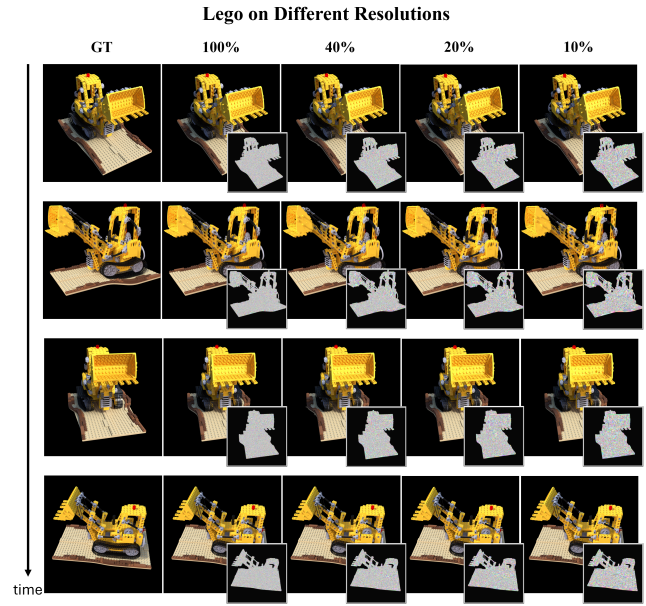

Fig. 9. The Lego animated with animation in different resolutions.

[32] Shenhan Qian, Tobias Kirschstein, Liam Schoneveld, Davide Davoli, Simon Giebenhain, and Matthias Nießner. 2023. GaussianAvatars: Photorealistic Head Avatars with Rigged 3D Gaussians. *arXiv preprint arXiv:2312.02069* (2023).
[33] Jiawei Ren, Liang Pan, Jiaxiang Tang, Chi Zhang, Ang Cao, Gang Zeng, and Ziwei Liu. 2023. DreamGaussian4D: Generative 4D Gaussian Splatting. *arXiv preprint arXiv:xxxx.xxxxx* (2023).
[34] Robin Rombach, Andreas Blattmann, Dominik Lorenz, Patrick Esser, and Björn Ommer. 2022. In *Proceedings of the IEEE/CVF conference on computer vision and pattern recognition*. 10684–10695.
[35] Patrick Schmidt, Dörte Pieper, and Leif Kobbelt. 2023. Surface Maps via Adaptive Triangulations. In *Computer Graphics Forum*, Vol. 42. Wiley Online Library, 103–117.
[36] Johannes L Schonberger and Jan-Michael Frahm. 2016. Structure-from-motion revisited. In *Proceedings of the IEEE conference on computer vision and pattern recognition*. 4104–4113.
[37] Zackary PT Sin, Peter HF Ng, and Hong Va Leong. 2023. NeRFahedron: A Primitive for Animatable Neural Rendering with Interactive Speed. *Proceedings of the ACM on Computer Graphics and Interactive Techniques* 6, 1 (2023), 1–20.
[38] Jiaxiang Tang, Jiawei Ren, Hang Zhou, Ziwei Liu, and Gang Zeng. 2023. DreamGaussian: Generative Gaussian Splatting for Efficient 3D Content Creation. *arXiv preprint arXiv:2309.16653* (2023).
[39] Jiaxiang Tang, Hang Zhou, Xiaokang Chen, Tianshu Hu, Errui Ding, Jingdong Wang, and Gang Zeng. 2023. Delicate textured mesh recovery from nerf via adaptive surface refinement. *arXiv preprint arXiv:2303.02091* (2023).
[40] Bugra Tekin, Federica Bogo, and Marc Pollefeys. 2019. H+O: Unified egocentric recognition of 3d hand-object poses and interactions. In *Proceedings of the IEEE/CVF conference on computer vision and pattern recognition*. 4511–4520.
[41] Ayush Tewari, Ohad Fried, Justus Thies, Vincent Sitzmann, Stephen Lombardi, Kalyan Sunkavalli, Ricardo Martin-Brualla, Tomas Simon, Jason Saragih, Matthias Nießner, et al. 2020. State of the art on neural rendering. In *Computer Graphics Forum*, Vol. 39. Wiley Online Library, 701–727.
[42] Justus Thies, Michael Zollhöfer, and Matthias Nießner. 2019. Deferred neural rendering: Image synthesis using neural textures. *Acm Transactions on Graphics (TOG)* 38, 4 (2019), 1–12.
[43] Alexander Vilesov, Pradyumna Chari, and Achuta Kadambi. 2023. CG3D: Compositional Generation for Text-to-3D via Gaussian Splatting. *Arxiv* (2023).
[44] Yiming Wang, Qin Han, Marc Habermann, Kostas Daniilidis, Christian Theobalt, and Lingjie Liu. 2023. NeuS2: Fast Learning of Neural Implicit Surfaces for Multi-view Reconstruction. In *Proceedings of the IEEE/CVF International Conference on Computer Vision (ICCV)*.

[45] Guanjun Wu, Taoran Yi, Jiemin Fang, Lingxi Xie, Xiaopeng Zhang, Wei Wei, Wenyu Liu, Qi Tian, and Xinggang Wang. 2023. 4d gaussian splatting for real-time dynamic scene rendering. *arXiv preprint arXiv:2310.08528* (2023).
[46] Wenqi Xian, Jia-Bin Huang, Johannes Kopf, and Changil Kim. 2021. Space-time Neural Irradiance Fields for Free-Viewpoint Video. 9421–9431.
[47] Ziyi Yang, Xinyu Gao, Wen Zhou, Shaohui Jiao, Yuqing Zhang, and Xiaogang Jin. 2023. Deformable 3d gaussians for high-fidelity monocular dynamic scene reconstruction. *arXiv preprint arXiv:2309.13101* (2023).
[48] Zeyu Yang, Hongye Yang, Zijie Pan, Xiatian Zhu, and Li Zhang. 2023. Real-time photorealistic dynamic scene representation and rendering with 4d gaussian splatting. *arXiv preprint arXiv:2310.10642* (2023).
[49] Lior Yariv, Peter Hedman, Christian Reiser, Dor Verbin, Pratul P Srinivasan, Richard Szeliski, Jonathan T Barron, and Ben Mildenhall. 2023. BakedSDF: Meshing Neural SDFs for Real-Time View Synthesis. *arXiv preprint arXiv:2302.14859* (2023).
[50] Taoran Yi, Jiemin Fang, Guanjun Wu, Lingxi Xie, Xiaopeng Zhang, Wenyu Liu, Qi Tian, and Xinggang Wang. 2023. Gaussiandreamer: Fast generation from text to 3d gaussian splatting with point cloud priors. *arXiv preprint arXiv:2310.08529* (2023).
[51] Heng Yu, Joel Julin, Zoltán Á Milacski, Koichiro Niinuma, and László A Jeni. 2023. CoGS: Controllable Gaussian Splatting. *arXiv preprint arXiv:2312.05664* (2023).
[52] Wentao Yuan, Zhaoyang Lv, Tanner Schmidt, and Steven Lovegrove. 2021. STaR: Self-supervised Tracking and Reconstruction of Rigid Objects in Motion with Neural Rendering. *arXiv preprint arXiv:2101.01602* (2021).
[53] Junwu Zhang, Zhenyu Tang, Yatian Pang, Xinhua Cheng, Peng Jin, Yida Wei, Wangbo Yu, Munan Ning, and Li Yuan. 2023. Repaint123: Fast and High-quality One Image to 3D Generation with Progressive Controllable 2D Repainting. arXiv:2312.13271 [cs.CV]
[54] Richard Zhang, Phillip Isola, Alexei A Efros, Eli Shechtman, and Oliver Wang. 2018. The unreasonable effectiveness of deep features as a perceptual metric. In *Proceedings of the IEEE conference on computer vision and pattern recognition*. 586–595.
[55] Wojciech Zielonka, Timur Bagautdinov, Shunsuke Saito, Michael Zollhöfer, Justus Thies, and Javier Romero. 2023. Drivable 3D Gaussian Avatars. (2023). arXiv:2311.08581 [cs.CV]
[56] Matthias Zwicker, Hanspeter Pfister, Jeroen Van Baar, and Markus Gross. 2002. EWA splatting. *IEEE Transactions on Visualization and Computer Graphics* 8, 3 (2002), 223–238.